\documentclass[12pt]{article}
\usepackage{amssymb,amsmath,bm}
\usepackage{fullpage}
\usepackage{mathtools}
\usepackage[round]{natbib}
\usepackage[colorlinks=true,linkcolor=black,citecolor=black,urlcolor=blue,pdfborder={0 0 0}]{hyperref}
\usepackage{tikz}
\usepackage{pgfplots}

\begin{document}
\title{Comment on `Linear energy transfer incorporated intensity modulated proton therapy optimization'}
\author{Bram L. Gorissen \\ \small{Department of Radiation Oncology, Massachusetts General Hospital and Harvard Medical School}\\
		\small{Boston, Massachusetts}}
\date{}
\maketitle
\begin{abstract}
	Cao et al (2018) published an article on inverse planning based on dose-averaged linear energy transfer (LET). Their claim that the problem can be cast as a linear optimization model relies on an incorrect application of the Charnes-Cooper transformation. In this comment we show that their linear model is simlar to one where dose-averaged LET is multiplied with dose, explaining why their model was nonetheless able to improve the LET distribution.
\end{abstract}
\begin{tikzpicture}[remember picture,overlay]
\node[anchor=south,yshift=10pt] at (current page.south) {\fbox{\parbox{\dimexpr\textwidth-\fboxsep-\fboxrule\relax}{\footnotesize This is an author-created, un-copyedited version of an article published in Physics in Medicine and Biology \href{https://doi.org/10.1088/1361-6560/aaffa6}{DOI:10.1088/1361-6560/aaffa6}.}}};
\end{tikzpicture}

With great interest we read the recently published article by \citet{cao2018linear} on linear energy transfer (LET) based inverse planning for intensity-modulated proton therapy (IMPT). The biological effectiveness of IMPT is at least correlated with LET \citep{peeler2016clinical}, and therefore, it is desirable to take LET into account when designing a treatment plan. Cao et al.~proposed a linear optimization model that minimizes LET in organs at risk (OARs) and maximizes LET in the target. However, the derivation of their model is erroneous due to which the solutions are suboptimal with respect to dose and LET.

The authors claim to optimize a plan based on the following objective function:
\begin{align}
f(\bm{w}) = \; &\frac{\lambda_T^+}{|T|} \sum_{i \in T} \max\left\{ 0, d_i - d_i^{\text{pr}}\right\} + \frac{\lambda_T^-}{|T|} \sum_{i \in T} \max\left\{ 0, d_i^{\text{pr}} - d_i\right\} \label{l1} \\
& \quad + \frac{\lambda_O}{|O|} \sum_{i \in O} \max\left\{ 0, d_i - d_i^{\text{max}}\right\} + \frac{\lambda_N}{|N|} \sum_{i \in N} d_i \label{l2} \\
&\quad - \frac{\theta_T}{|T|} \sum_{i \in T} l_i + \frac{\theta_O}{|O|} \sum_{i \in O} l_i, \label{l3}
\end{align}
where the dose in voxel $i$ is given by $d_i = \sum_j d_{ij} w_j$ with $w_j$ the weight of pencil beam $j$, and $l_i$ is the dose-averaged LET $\sum_j D_{ij} L_{ij} w_j / \sum_j D_{ij} w_j$. The first two lines of this objective function are well known in inverse planning: the terms in \eqref{l1} incur a penalty when the dose in target $T$ is below or above the prescribed dose $d_i^{\text{pr}}$, while the terms in \eqref{l2} penalize dose above the maximum prescribed dose $d_i^{\text{max}}$ in OARs $O$ and total dose in normal tissue $N$. The novel part are the terms in \eqref{l3} that maximize dose averaged LET in the target and minimize dose averaged LET in the OARs. The weights $\lambda$ and $\theta$ control the trade-off between the different objectives, and were tuned for each patient to yield similar physical dose as previous clinical plans.

It is well known that the terms in \eqref{l1} and \eqref{l2} can be linearized and that it is easy to find a global optimum for just these terms. It is considerably harder to optimize \eqref{l3}, which is a sum of fractions of linear functions after substituting the formula for $l_i$. \cite{cao2018linear} claim that the variable transformation proposed by \citet{charnes1962programming} results in an equivalent formulation that is linear. {\it This cannot be true, because this transformation can be applied to a single fraction only}, not to a sum of fractions unless all denominators are the same. Although it is hard to prove that a linear reformulation of \eqref{l3} does not exist, it is clear that \citet{cao2018linear} have used a method that is not suitable.

What we could ask ourselves is what Cao et al.~have actually optimized, and if there is perhaps an explanation as to why their optimization problem finds treatment plans with an improved LET distribution. It is impossible to answer this question with certainty based on the information reported in the paper. The authors mention a substitution $t=1/d_i$, but since the left hand side does not depend on $i$, it is not clear which voxel was used to define $t$. If we ignore $t$, the final objective used by Cao et al.~is:
\begin{align}
g(\bm{x}) = \; &\frac{\lambda_T^+}{|T|} \sum_{i \in T} \max\left\{ 0, d_i - d_i^{\text{pr}}\right\} + \frac{\lambda_T^-}{|T|} \sum_{i \in T} \max\left\{ 0, d_i^{\text{pr}} - d_i\right\}  \notag \\
& \quad + \frac{\lambda_O}{|O|} \sum_{i \in O} \max\left\{ 0, d_i - d_i^{\text{max}}\right\} + \frac{\lambda_N}{|N|} \sum_{i \in N} d_i \notag \\
&\quad - \frac{\theta_T}{|T|} \sum_{i \in T} (ld)_i + \frac{\theta_O}{|O|} \sum_{i \in O} (ld)_i, \label{eq:ld}
\end{align}
where $(ld)_i$ is the dose-averaged LET multiplied with dose: $\sum_j D_{ij} L_{ij} x_j$. Optimization problems with $(ld)_i$ have been proposed as a first approximation to the additional biological dose due to high LET \citep{unkelbach2016reoptimization}. Cao et al.~seemingly missed this interpretation of \eqref{eq:ld}, and did not directly use the pencil beam weights $\bm{x}$ that are optimal to the objective function \eqref{eq:ld}. Instead, they took the optimal $\bm{x}$, and a posteriori scaled it with $t$ to ensure constraint satisfaction. This step is rather unnecessary since the constraints (omitted here for brevity) could have been formulated in a way that an a posteriori scaling is not needed. Our expectation is that $t$ was sufficiently close to 1 for all patients, so that the presented scaled treatment plans were near-optimal to \eqref{eq:ld}.

Everything considered, the results by Cao et al.~are not as groundbreaking as they may seem. In contrast to what the authors claim, they have not directly optimized dose averaged LET, but instead used a model where dose averaged LET was multiplied with dose. Models with such terms are well known and have a meaningful interpretation, which the authors seemingly missed. Due to the added scaling factor $t$, the interpretability and optimality of their solution is impaired.

\section*{Acknowledgment}
Supported in part by NIH U19 Grant 5U19CA021239-38.

\bibliographystyle{abbrvnat}
\bibliography{bibfile}

\begin{thebibliography}{4}
\providecommand{\natexlab}[1]{#1}
\providecommand{\url}[1]{\texttt{#1}}
\expandafter\ifx\csname urlstyle\endcsname\relax
  \providecommand{\doi}[1]{doi: #1}\else
  \providecommand{\doi}{doi: \begingroup \urlstyle{rm}\Url}\fi

\bibitem[Cao et~al.(2018)Cao, Khabazian, Yepes, Lim, Poenisch, Grosshans, and
  Mohan]{cao2018linear}
W.~Cao, A.~Khabazian, P.~P. Yepes, G.~Lim, F.~Poenisch, D.~R. Grosshans, and
  R.~Mohan.
\newblock Linear energy transfer incorporated intensity modulated proton
  therapy optimization.
\newblock \emph{Physics in Medicine \& Biology}, 63\penalty0 (1):\penalty0
  015013, 2018.

\bibitem[Charnes and Cooper(1962)]{charnes1962programming}
A.~Charnes and W.~W. Cooper.
\newblock Programming with linear fractional functionals.
\newblock \emph{Naval Research logistics quarterly}, 9\penalty0 (3-4):\penalty0
  181--186, 1962.

\bibitem[Peeler et~al.(2016)Peeler, Mirkovic, Titt, Blanchard, Gunther,
  Mahajan, Mohan, and Grosshans]{peeler2016clinical}
C.~R. Peeler, D.~Mirkovic, U.~Titt, P.~Blanchard, J.~R. Gunther, A.~Mahajan,
  R.~Mohan, and D.~R. Grosshans.
\newblock Clinical evidence of variable proton biological effectiveness in
  pediatric patients treated for ependymoma.
\newblock \emph{Radiotherapy and Oncology}, 121\penalty0 (3):\penalty0
  395--401, 2016.

\bibitem[Unkelbach et~al.(2016)Unkelbach, Botas, Giantsoudi, Gorissen, and
  Paganetti]{unkelbach2016reoptimization}
J.~Unkelbach, P.~Botas, D.~Giantsoudi, B.~L. Gorissen, and H.~Paganetti.
\newblock Reoptimization of intensity modulated proton therapy plans based on
  linear energy transfer.
\newblock \emph{International Journal of Radiation Oncology* Biology* Physics},
  96\penalty0 (5):\penalty0 1097--1106, 2016.

\end{thebibliography}
\end{document}